# Inter-layer Potential for Hexagonal Boron Nitride

*Itai Leven,[1] Ido Azuri,[2] Leeor Kronik,[2] and Oded Hod[1,*]*

[1] Department of Chemical Physics, School of Chemistry, the Raymond and Beverly Sackler Faculty of Exact Sciences, Tel-Aviv University, Tel-Aviv 69978, Israel

[2] Department of Materials and Interfaces, Weizmann Institute of Science, Rehovoth 76100, Israel

ABSTRACT:

A new interlayer force-field for layered hexagonal boron nitride ($h$-BN) based structures is presented. The force-field contains three terms representing the interlayer attraction due to dispersive interactions, repulsion due to anisotropic overlaps of electron clouds, and monopolar electrostatic interactions. With appropriate parameterization, the potential is able to simultaneously capture well the binding and lateral sliding energies of planar $h$-BN based dimer systems as well as the interlayer telescoping and rotation of double walled boron-nitride nanotubes of different crystallographic orientations. The new potential thus enables the accurate and efficient modeling and simulation of large-scale $h$-BN based layered structures.

[*]Corresponding author: odedhod@tau.ac.il

# 1. Introduction

The fabrication of novel nano-scale materials has led to new opportunities in a wide range of technological fields, one of which focuses on their utilization in miniaturized mechanical systems. Systems such as nano-motors, nano-bearings, and nano-resonators have been demonstrated in both theory and experiment to hold great potential for use as components in future nanoelectromechanical devices.[1-5] Originally, the main focus has been on carbon based materials, such as graphene[6] and carbon nanotubes[7] (CNTs) due to their extraordinary mechanical rigidity[8] and tunable electronic properties.[9,10] Recently, the technological potential of other layered materials, such as hexagonal boron nitride ($h$-BN) and its derivative structures, has been realized. $h$-BN substrates have been shown to enhance the performance of graphene based electronic devices[2,6,7,11-18] and the heterojunction constructed from these two materials has been predicted to present robust superlubric behavior.[19] Furthermore, boron nitride nanotubes (BNNTs) have been shown to serve as a better hydrogen storage medium than CNTs.[20,21] and their torsional stiffness has been shown[18] to be up to 10 times higher than that of CNTs.[22]

$h$-BN belongs to the same family of stacked hexagonal materials as graphene, which consists of atomically thin sheets held together by weak van der Waals (vdW) forces.[23] The intra-layer network of $h$-BN consists of strong $sp^2$ covalent bonds, whose partial ionic character turn the system into an insulator.[24] Similar to graphene, the anisotropic binding of $h$-BN allows for the formation of various layered structures, such as tubes, sheets, cones, and scrolls. Long-range interlayer interactions play a crucial role in characterizing the structural, mechanical, and tribological properties of these systems and hence their performance in nanoelectromechanical devices.[1,5,22,25-27]

In order to model such systems, several electronic structure methods, that aim to provide a balanced description of covalent, ionic, and long-range dispersive interactions, have been developed.[15,28,29] These methods are most suitable for treating relatively small molecular systems and crystalline materials characterized by compact unit-cells. When studying the dynamical tribological properties of large-scale material interfaces, one often resorts to molecular dynamics (MD) simulations based on appropriately parameterized classical force-fields. Here, it was shown that traditional force-field potentials like the Lennard-Jones potential may produce quantitative and even qualitative disagreement with results obtained via more accurate computational methods.[25,27,30] Hence, carefully tailored force-fields have to be developed in order to obtain physically meaningful results.[16] An example for such an interlayer force-field is the registry-dependent potential developed by Kolmogorov and Crespi for graphitic systems.[31] This potential, which takes into account the anisotropic nature of the inter-layer interactions, has been shown to provide an appropriate balance between accuracy and computational burden in MD simulations of double-walled carbon nanotubes and layered graphene structures.[4,32,33]

In the present paper, we present a new interlayer force-field for *h*-BN, which captures all important physical parameters required to describe its interlayer interactions. Our method, termed *h*-BN ILP (inter-layer potential), incorporates an attractive component adopted from the Tkatchenko-Scheffler dispersive correction (TS-vdW)[34] to density functional theory (DFT), a repulsive term based on the graphitic registry dependent inter-layer potential of Kolmogorov and Crespi,[31] and a classical monopolar electrostatic term that takes into account the partially ionic character of *h*-BN. It is shown that, by appropriate parameterization against a series of benchmark calculations performed using advanced DFT calculations, the new force-field is able to accurately describe both the interlayer *binding* and *sliding* energy landscapes of complex materials junctions based on *h*-BN.

The paper is organized as follows: in the next section we provide a detailed account of the form of the proposed force-field. This is followed by a description of benchmark calculations performed for the parameterization of the potential. Next, we validate the method by comparing the *h*-BN ILP calculated interlayer telescoping and rotation of double-walled boron-nitride nanotubes (DWBNNTs) to results obtained via DFT calculations. Finally, we conclude and discuss future applications.

## 2. Force-Field description

The *h*-BN ILP force-field is constructed to exclusively treat the *interlayer* interactions in layered structures based on *h*-BN. It is designed to augment existing force-fields, which already provide a good treatment of intra-layer interactions but fail in providing an appropriate description of the interlayer physics. Here, for describing the intra-layer interactions, we use ReaxFF, designed by Goddard *et al*. and shown to yield good results for simulating bond breakage and formation for a wide variety of chemical systems.[35] We somewhat modify the parameterization of ReaxFF for B-N-H molecules, originally calibrated for ammonia borane dehydrogenation and combustion,[36] so as to improve specifically its performance for *h*-BN and its derivatives. The new parameterization is fine-tuned against a set of benchmark calculations of bond dissociation, valence angles, and torsional angle strain calculated for several B-N-H molecular derivatives. For consistency with previous work, these calculations have been performed at the B3LYP/6-31G** level of DFT (see supplementary material).[35,37-39] We note that since, by constructions, the inter- and intra-layer interactions are treated separately, *h*-BN ILP requires (*a priori*) knowledge regarding the layer attribution of each atom in the system.

The *h*-BN ILP consists of three terms which describe the long-range dispersive interactions, repulsions between overlapping π electron clouds, and monopolar electrostatic interactions between the different layers. The latter term stems from the polar nature of the intra-layer covalent B-N bonds. In the following, we provide a detailed description of the functional form of these three terms and numerical values of the various parameters that the *h*-BN ILP uses.

a. <u>Dispersive term</u>

To treat long-range attractive van-der Waals interactions, we adopt a dispersion correction similar to the one developed by Tkatchenko and Scheffler to augment standard ground state DFT exchange-correlation functional approximations.[34] This term incorporates a $C_6/r^6$ Lennard-Jones type interaction, which is damped in the short-range to avoid double counting of correlation effects :

$$E_{Dis}(r_{ij}) = \text{Tap}(r_{ij}) \left\{ -\frac{1}{1+e^{-d\left[\left(r_{ij}/\left(s_R \cdot r_{ij}^{eff}\right)\right)-1\right]}} \cdot \frac{C_{6,ij}}{r_{ij}^6} \right\} \quad (1)$$

Here, $r_{ij}$ is the distance between atoms *i* and *j* located on different layers, $d$ and $s_R$ are unit-less parameters dictating the steepness and onset of the short-range Fermi-type damping function, and $r_{ij}^{eff}$ and $C_{6,ij}$ are the effective atomic radii and pair-wise dispersion coefficients, in the molecular environment, respectively. The original TS expression, which consists of the term in curly brackets, is further augmented by the taper correction,[40] $\text{Tap}(r_{ij})$, which provides a continuous (up to 3rd derivative) long-range cutoff term of the form $\text{Tap}(r_{ij}) = \frac{20}{R_{cut}^7} r_{ij}^7 - \frac{70}{R_{cut}^6} r_{ij}^6 + \frac{84}{R_{cut}^5} r_{ij}^5 - \frac{35}{R_{cut}^4} r_{ij}^4 + 1$ for inter-atomic separations larger than $R_{cut}$. This cutoff is often employed in calculations of large systems in order to reduce the computational burden by minimizing the number of atomic pairs that are taken into account. We note that care should be taken when using such cutoffs in periodic systems, where the infinite sums may become conditionally convergent.[23]

In the TS-vdW correction scheme, the sum of effective atomic radii and the dispersion coefficients are evaluated from the ground-state electron density at each molecular configuration. For the $h$-BN systems studied, we found a variation of ~10% in the value of these parameters throughout the TS-DFT benchmark calculations that we have performed. Therefore, fixed values for these parameters were chosen. We note that in order to obtain an appropriate balance between the three long-range terms of the $h$-BN ILP, $C_6$ coefficients larger by ~20% than those obtained via the TS-vdW approach had to be chosen. We attribute this discrepancy to the fact that the complex quantum/classical electrostatic interactions appearing in the first-principles calculations are replaced by a simple classical Coulombic monopolar term in the force-field. Hence, within our efforts of maintaining the intra-layer force-field unchanged, we are forced to adjust the interlayer terms to compensate for this simplification. Nevertheless, as shown below, our choice of dispersion coefficients allows for a good agreement between the overall binding and sliding energy landscapes of various $h$-BN structures calculated via advanced first-principles methods and via the $h$-BN ILP.

b. <u>Repulsive term</u>

The repulsive term of the $h$-BN ILP accounts for Pauli repulsions between overlapping electron clouds around atoms belonging to different layers. In order to appropriately describe the anisotropy of the $\pi$ electron cloud around the boron and nitrogen atomic sites, the repulsive term requires information regarding both the actual ($r_{ij}$) and the lateral ($\rho_{ij}$) distance (given in units of Å throughout the manuscript) between the $i$ and $j$ atomic sites (see Fig. 1). Here, the lateral distance $\rho_{ij}$ is defined as the distance between atom $j$ of one layer and the normal associated with atom $i$ residing on the adjacent layer. The normal to atom $i$, in turn, is defined as the line perpendicular to the plane formed by the three nearest bonded neighbors while passing through atom $i$ itself (see Fig. 1a). Note that according to this definition, apart from the case of perfectly parallel layers, in general $\rho_{ij} \neq \rho_{ji}$.

With this definition, the repulsive term is constructed from a Morse-like exponential isotropic term, multiplied by an anisotropic correction, which enhances the repulsion at small inter-atomic lateral separations in the following manner:[31]

$$E_{Rep} = \text{Tap}(r_{ij})e^{\alpha_{ij}\cdot\left(1-\frac{r_{ij}}{\beta_{ij}}\right)}\left[\varepsilon_{ij} + C\left(e^{-\left(\frac{\rho_{ij}}{\gamma_{ij}}\right)^2} + e^{-\left(\frac{\rho_{ji}}{\gamma_{ij}}\right)^2}\right)\right] \quad (2)$$

Here, $\text{Tap}(r_{ij})$ is the above-discussed tapering long-range cut-off scheme, $\alpha_{ij}$ and $\beta_{ij}$ set the slope and range of the isotropic Pauli repulsive wall, and $\gamma_{ij}$ sets the width of the Gaussian decay factors in the anisotropic correction term and thus dictates the sensitivity of the repulsive term to the lateral distance between atoms $i$ and $j$. $C$ and $\varepsilon_{ij}$ are constant scaling factors bearing units of energy.[41]

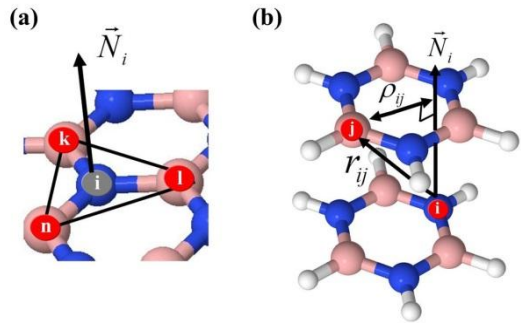

Figure 1: *Definitions of the lateral distance and normal vector: (a) The normal of atom i is defined as the normal of the plane formed by its three nearest directly-bonded neighbors k, l and n. (b) The lateral distance, $\rho_{ij}$, is defined as the distance between atom j and the normal of atom i.*

c. Electrostatic term

To take interlayer monopolar electrostatic interactions into account we adopt the formalism implemented in the ReaxFF scheme.[35] Within this approach, a Coulomb potential energy term of the form $E_{Coulomb}(r_{ij}) = Tap(r_{ij}) \cdot 23.0609 \cdot \kappa\, q_i q_j / \sqrt[3]{r_{ij}^3 + (1/\lambda_{ij})^3}$ is used to describe the classical monopolar electrostatic potential energy between partially charged ionic lattice sites $i$ and $j$ residing on different layers. Here, $Tap(r_{ij})$ is the tapering long-range cut-off scheme discussed above, $\kappa$ is Coulomb's constant in units of eV·Å (see table of parameters below), and $\lambda_{ij} = \sqrt{\lambda_{ii} \cdot \lambda_{jj}}$ is a shielding parameter eliminating the short-range singularity of the classical monopolar electrostatic interactions in regions where repulsive interactions between overlapping electron clouds dominate the interlayer potential. The factor 23.0609 kcal/mol·eV converts the units of energy from eV to kcal/mol.

The partial ionic charges, $q_i$ and $q_j$ (given in units of the elementary charge throughout the manuscript), which are associated with atomic sites on the different layers, are calculated using the electronegativity equalization method (EEM).[35,42-45] This method is based on a principle devised by Sanderson[46,47] stating that, upon molecular formation, the electronegativities of the constituent atoms equalize to give a global molecular electronegativity. Here, the electronegativity of a given atom within the molecular environment ($\chi_i$) is written in terms of the corresponding isolated atom electronegativity[48] ($\chi_i^0$) and hardness[49] ($\eta_i^0$) in the following manner:

$$\chi_i = (\chi_i^0 + \Delta\chi_i) + 2(\eta_i^0 + \Delta\eta_i)q_i + \sum_{j \neq i}^{N} \kappa q_j / \sqrt[3]{r_{ij}^3 + (1/\lambda_{ij})^3} \qquad (3)$$

Where $\Delta\chi_i$ and $\Delta\eta_i$ account for the variation of the isolated atomic electronegativity and hardness, due to the molecular environment, and $q_j$ is the effective atomic charge.

The effect of the molecular environment, on the electronegativity of atom *i* is taken into account via the last term, representing an external shielded-coulomb potential due to the effective atomic charges associated with the remaining *N*-1 atoms in the molecule.

The effective atomic charges are obtained by a set of equations forcing all atomic electronegativities be equal to the (generally unknown) equilibrated molecular electronegativity $(\chi_{eq} = \chi_1 = \chi_2 = \cdots = \chi_N)$ and an extra equation demanding that the sum of atomic charges equal the total molecular charge, *Q*. These equations are given in matrix form as follows:

$$\begin{pmatrix} 2\eta_1^* & \kappa/R_{12} & \cdots & \kappa/R_{1N} & -1 \\ \kappa/R_{21} & 2\eta_2^* & \cdots & \kappa/R_{2N} & -1 \\ \vdots & \vdots & \ddots & \vdots & \vdots \\ \kappa/R_{N1} & \kappa/R_{N2} & \cdots & 2\eta_N^* & -1 \\ 1 & 1 & \cdots & 1 & 0 \end{pmatrix} \begin{pmatrix} q_1 \\ q_2 \\ \vdots \\ q_N \\ \chi_{eq} \end{pmatrix} = - \begin{pmatrix} \chi_1^* \\ \chi_2^* \\ \vdots \\ \chi_N^* \\ -Q \end{pmatrix} \quad (4)$$

where we have defined $\chi_i^* \equiv \chi_i^0 + \Delta\chi_i$ and $\eta_i^* \equiv \eta_i^0 + \Delta\eta_i$. Given a complete set of appropriately parameterized values of the electronegativities $\{\chi_i^*\}$ and hardnesses $\{\eta_i^*\}$ one can solve these equations to obtain the effective atomic charges. To this end, the free-atom electronegativities $\{\chi_i^0\}$ and hardnesses $\{\eta_i^0\}$ can be extracted from available literature[49,50] and the corresponding sets of corrections $\{\Delta\chi_i\}$ and $\{\Delta\eta_i\}$ can be calibrated against *ab-initio* calculations for a benchmark set of molecules. In the present work, we start with the set of $\{\chi_i^*\}$ and $\{\eta_i^*\}$ values suggested by Weismiller *et al.* for ammonia-borane derivatives.[36] and adjust them to obtain good correspondence between the calculated atomic charges and Mulliken atomic charges obtained for several *h*-BN molecular derivatives using the B3LYP exchange-correlation functional approximation[37,51] of density functional theory with the 6-31G** basis set, as implemented in the Gaussian suite of programs.[38,52] A comparison of the effective atomic charges, calculated using EEM with our choice of parameters, to the DFT results for a set of B-N-H molecules can be found in the supplementary material.

d. Complete list of parameters and their numerical values

The following table summarizes all parameters appearing in the *h*-BN ILP and their recommended numerical values as used in the calculations below.

| Term | Parameter | Value | | | | | | Units |
|---|---|---|---|---|---|---|---|---|
| | | BB | NN | HH | BN | BH | NH | |
| Dispersive | $d$ | 15.0 | | | | | | |
| | $S_r$ | 0.84 | | | | | | |
| | $r_{ij}^{eff}$ | 3.77 | 3.361 | 2.797 | 3.566 | 3.284 | 3.08 | Å |
| | $C_{6,ij}$ | 968.448 | 295.351 | 41.231 | 505.965 | 196.76 | 108.956 | $kcal \cdot Å^6/mol$ |
| Taper | $R_{cut}$ | 20.0 | | | | | | Å |
| Repulsive | $\alpha_{ij}$ | 11.0 | 12.08 | 9.0 | 10.7 | 9.0 | 9.0 | |
| | $\beta_{ij}$ | 3.1 | 3.69 | 2.7 | 2.6 | 2.8 | 2.7 | Å |
| | $\gamma_{ij}$ | 1.8 | 1.2 | 20.0 | 1.8 | 20.0 | 20.0 | Å |
| | $\varepsilon_{ij}$ | 0.46 | 0.21 | 0.31 | 0.2 | 0.31 | 0.25 | $kcal/mol$ |
| | $C$ | 0.068 | | | | | | $kcal/mol$ |
| Electrostatic | $\kappa$ | 14.4 | | | | | | $eV \cdot Å$ |
| | $\lambda_{ij}$ | 0.7 | 0.69 | 0.8 | | | | $Å^{-1}$ |
| EEM | $\chi_i^*$ | *10.0* | 13.0 | 10.2 | | | | eV |
| | $\eta_i^*$ | 6.702 | 7.0 | 7.0327 | | | | eV |
| | $\lambda_{ij}$ | 0.7 | 0.69 | 0.8 | | | | $Å^{-1}$ |

3. **Benchmark calculations**

The *h*-BN ILP parameters presented above have been calibrated via a set of benchmark DFT calculations of binding and sliding energies of several finite, hydrogen passivated, *h*-BN flake dimers. For the binding energy calculations, geometry optimization at the HSE/6-31G** [53-55] DFT level of theory has been performed on individual monomers of increasing size, which were then duplicated to form the dimers. Binding energy curves of the rigid flakes with varying inter-flake distance were then obtained using the B3LYP exchange-correlation density functional approximation augmented with the TS-vdW correction.[56,57] We note that the HSE and B3LYP geometries are practically identical (with BN bond length variations of 0.01 Å) yielding binding energy curves which are almost indistinguishable. This indicates that our suggested parameterization is insensitive to the particular choice of underlying hybrid functional. For the smaller dimer systems considered (see insets in panels I-III of Fig. 2), these calculations have been performed using the full TS-vdW scheme as implemented in the FHI-AIMS code with the tier-2 basis set and tight convergence settings.[58,59] Here, the effects of basis set super position errors (BSSE) on the binding energy were found to be less than 7% and hence no BSSE correction was employed (see supporting information for further details). The computational burden of executing this procedure for the largest dimer system considered (see inset of Fig. 2 IV) was found to be too demanding. Therefore, the binding energy calculations for this dimer were performed in two stages: First, binding energy curves were obtained using the Gaussian suite of programs at the B3LYP/6-31G** level of theory utilizing the counterpoise BSSE correction.[52,60,61] Next, the TS-vdW correction was added according to Eq. (1) (while discarding the taper correction) and using fixed Hirshfeld atomic volumes[62] to set the effective atomic radii and pair-wise $C_6$ dispersion coefficients. Here, the fixed volumes were chosen as the average values extracted from the full TS-vdW binding energy calculations of the smaller dimer systems. These values were found to be quite robust, presenting variations of the order of 10%. More details regarding this calculation can be found in the supplementary material.

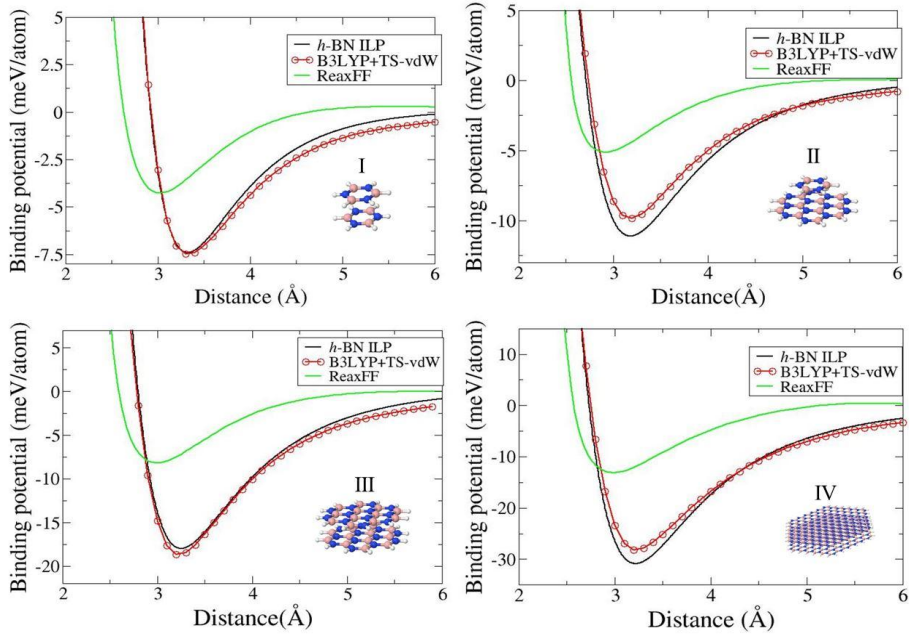

Figure 2: *Binding energy curves of the four dimer systems considered at the AA' stacking mode, calculated using the h-BN ILP (black line), B3LYP+TS-vdW (red line) and the ReaxFF as implemented within the LAMMPS code (green line). Insets: tilted views of the various dimer systems considered.*

In Fig. 2, we compare the binding energy curves calculated using the TS-vdW corrected B3LYP DFT calculation with results obtained using the *h*-BN ILP. For comparison purposes, we also present results of the original ReaxFF[35] scheme, as implemented in the LAMMPS[39,63] code and parameterized according to Ref. 36. As can be seen, for all dimers considered the *h*-BN ILP yields binding energy curves which are in very good agreement with the TS-vdW results.[64] Table 1 summarizes the binding energies and equilibrium inter-monomer distances as calculated by the B3LYP+TS-vdW, *h*-BN ILP, and ReaxFF methods. Here, the equilibrium binding distances of the B3LYP+TS-vdW calculations were extracted by fitting a 9$^{th}$ order polynomial to the calculated results. The largest deviation is obtained for the binding of borazine on a small *h*-BN flake (panel II of Fig. 2), where the *h*-BN ILP predicts a binding energy of -11.1 meV/atom, compared to -9.8 meV/atom evaluated with the B3LYP+TS-vdW method. The equilibrium distances calculated using the two methods in this case are in agreement to within 0.03 Å. For the largest dimer considered, consisting of 252 atoms, the *h*-BN ILP predicts a binding energy of -30.9 meV/atom, which is in good agreement with the B3LYP+TS-vdW result of -28.3 meV/atom. The inter-monomer equilibrium distances calculated for this system agree to within 0.02 Å. For all systems considered the ReaxFF, which is designed to capture the intra-layer chemistry but lacks appropriate treatment of interlayer interactions, considerably underestimates both the binding energies and the inter-monomer equilibrium distance. This exemplifies the importance of considering the anisotropic nature of the inter- and intra-layer interactions in *h*-BN based layered materials and the necessity for appropriate parameterization of the potential for describing these interactions simultaneously.

| Structure | I | | II | |
|---|---|---|---|---|
| | 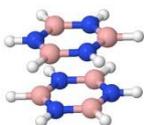 | | 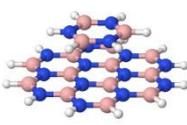 | |
| | Eq. distance (Å) | BE(meV/atom) | Eq. distance (Å) | BE(meV/atom) |
| B3LYP+TS-vdW | 3.33 | -7.5 | 3.2 | -9.8 |
| *h*-BN ILP | 3.32 | -7.4 | 3.2 | -11.1 |
| ReaxFF | 3.03 | -4.2 | 2.9 | -5.0 |
| Structure | III | | IV | |
| | 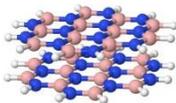 | | 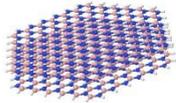 | |
| | Eq. distance (Å) | BE(meV/atom) | Eq. distance (Å) | BE(meV/atom) |
| B3LYP+TS-vdW | 3.24 | -18.7 | 3.2 | -28.3 |
| *h*-BN ILP | 3.25 | -18.0 | 3.2 | -30.9 |
| ReaxFF | 3.00 | -8.2 | 3.0 | -13.1 |

When considering the lateral interlayer sliding of hexagonal layered structures based on graphene and *h*-BN, vdW interactions have been shown to have a minor effect on the sliding energy corrugation.[25,27] For the system studied herein, the effect of including vdW interactions on the sliding energy surface corrugation is evaluated to be ~12.5% of the maximal corrugation. Despite its small relative contribution, this effect has been included in our sliding energy calculations using the TS-vdW correction with fixed Hirshfeld volumes following the procedure described above.

In Fig. 3, we present the sliding energy landscape of a borazine molecule on an *h*-BN sheet, represented by a finite flake constructed from 8x10 *h*-BN unit-cells, as obtained using DFT at the B3LYP+TSvdW/6-31G** level of theory, *h*-BN ILP, and the ReaxFF[35] as implemented in the LAMMPS[39,63] code and parameterized according to Ref. 36. The borazine molecule was initially placed 3.3 Å above the center of the larger *h*-BN flake, at the optimal AA' stacking mode where the boron (nitrogen) atoms of the borazine molecule reside atop nitrogen (boron) atoms of the sheet. The borazine molecule was then rigidly shifted in the lateral directions and single-point calculations were carried out at each shifted configuration until a sliding energy landscape representing a full unit-cell shift was achieved.

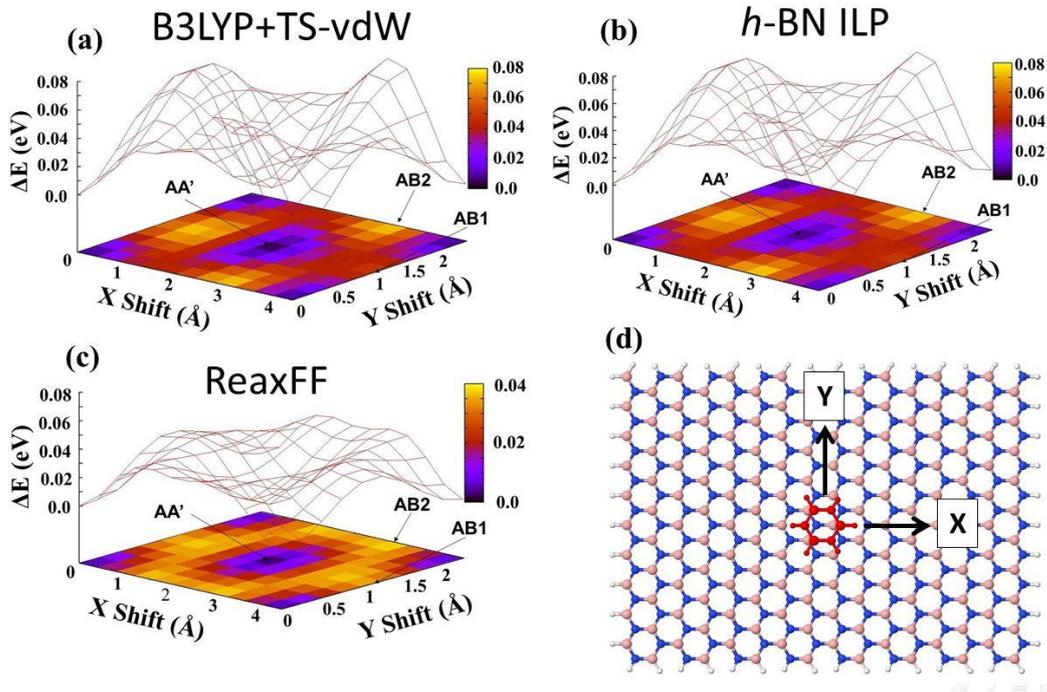

*Figure 3: Sliding energy landscape of a borazine molecule on an h-BN substrate, as calculated using the (a) B3LYP+TS-vdW/6-31G\*\*; (b) h-BN ILP; and (c) ReaxFF schemes. The energy of the AA' stacking mode is used as reference. Three high-symmetry stationary points are marked. A schematic representation of the molecular system is presented in panel (d), where the borazine molecule (marked in red) is placed on top of the edge-hydrogenated 8x10 h-BN rectangular flake representing the sheet.*

It is readily observed that the *h*-BN ILP, with the parameterization presented above, is able to capture not only the binding energy curves but also the lateral sliding energy landscape as calculated via dispersion-corrected DFT. As may be expected, the maximum repulsion is found at the $AB_2$ stacking mode, where the nitrogen atoms of the borazine molecule and the *h*-BN sheet are eclipsed.[27] The DFT and *h*-BN ILP interlayer potentials obtained at this position are 0.076 eV and 0.077 eV, respectively, relative to the total energy of the *AA'* configuration. The ReaxFF sliding energy landscape was found to deviate both quantitatively and qualitatively from the DFT results, producing an interlayer potential of 0.038 eV at the $AB_2$ stacking mode.

To exemplify further the performance of the $h$-BN ILP, we plot in Fig. 4 the maximum corrugation of the sliding energy landscape, defined as the energy difference between the worst ($AB_2$) and the optimal ($AA'$) stacking modes, as a function of the distance between the borazine molecule and the $h$-BN layer. Here as well, $h$-BN ILP reproduces well the DFT results, whereas the ReaxFF predicts a very weak dependence of the sliding energy corrugation on the molecule-sheet separation.

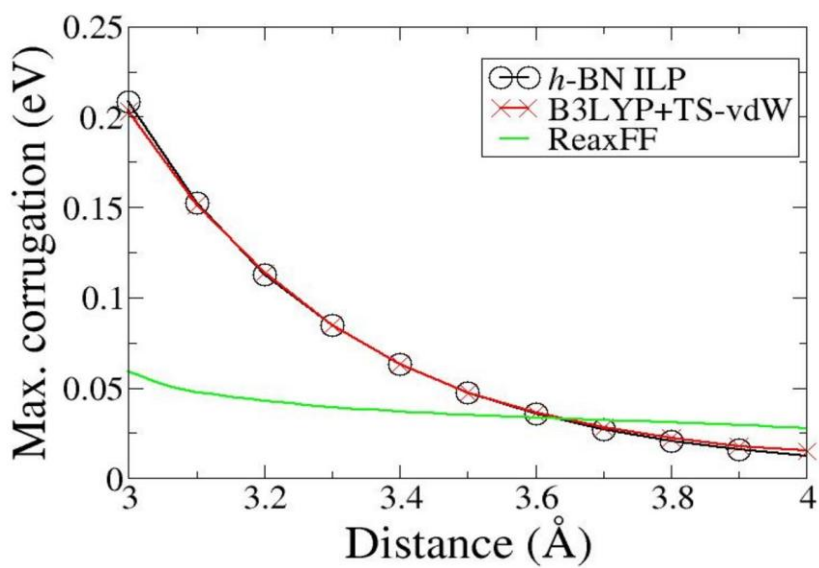

Figure 4: *Maximum corrugation of the sliding energy landscape of a borazine molecule on an h-BN sheet ($E[AB_2] - E[AA']$), as a function of molecule-sheet distance calculated by h-BN ILP (black), B3LYP+TS-vdW (red), and ReaxFF (green).*

## 4. Performance tests

As shown above, the parameterization of the $h$-BN ILP calibrated against planar dimer configurations produces excellent agreement with both binding and sliding vdW corrected DFT results. Here, we challenge the $h$-BN ILP against more complex curved structures. To this end, we consider two double-walled boron-nitride nanotubes: the armchair (5,5)@(10,10) and the zigzag (6,0)@(14,0), where $(n_1,m_1)@(n_2,m_2)$ designates a $(n_1,m_1)$ inner shell coaxially located within a $(n_2,m_2)$ outer shell. A single hydrogen passivated BNNT unit-cell is used to represent the outer shell, whereas the inner shell is represented by a five unit-cell long hydrogen passivated BNNT segment. A fixed BNNT geometry with B-N bond lengths of 1.44 Å, and B-H and N-H bond lengths of 1.18 Å and 1.0 Å, respectively, are chosen with no further geometry optimization. The outer ring, initially placed at the center of the inner tube, is rigidly axially shifted and rotated around the fixed inner shell. At each inter-tube configuration, a single-point calculation is carried out and the total energy is recorded.

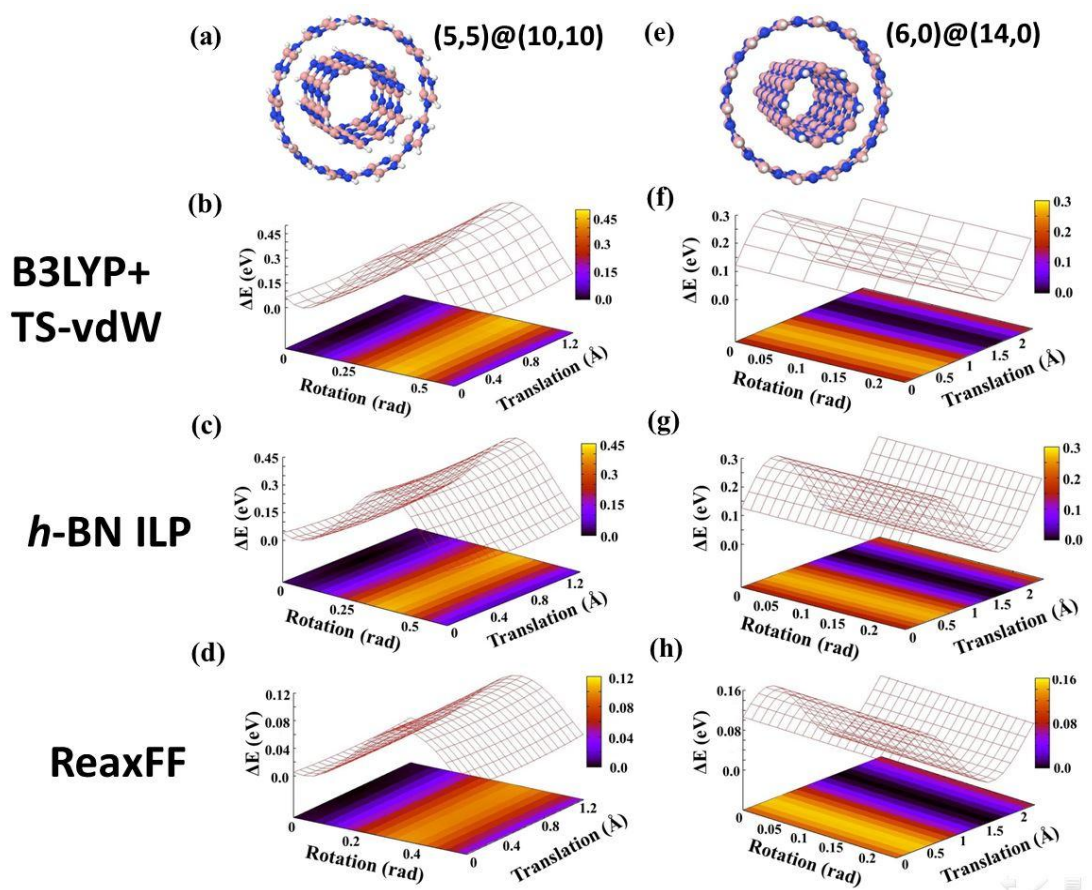

Figure 5: *Inter-tube sliding and rotation potential energy landscapes for a (5,5)@(10,10) (left panels) and a (6,0)@(14,0) (right panels) double-walled boron-nitride nanotubes. Presented are results of DFT calculations at the B3LYP+TSvdW/6-31G\*\* level of theory (second row), h-BN ILP (third row) and ReaxFF (last row). Illustrations of the model double-walled boron-nitride nanotubes used appear in the upper row.*

The resulting sliding-rotation energy surfaces are presented in Fig. 5. As in the case of the planar systems, the *h*-BN ILP captures well, both qualitatively and quantitatively, the vdW corrected DFT results. The maximum sliding-rotation corrugations obtained using our DFT and *h*-BN ILP calculations for the (5,5)@(10,10) DWBNNT are 0.46 eV and 0.41 eV, respectively. Similarly, for the (6,0)@(14,0) DWBNNT a value of 0.26 eV was obtained for both the DFT and *h*-BN ILP calculations. In the present case, the ReaxFF seems to produce qualitatively good results. Nevertheless, it fails to quantitatively reproduce the vdW corrected DFT results, yielding considerably lower corrugation values of 0.14 eV and 0.10 eV for the (6,0)@(14,0) and (5,5)@(10,10) DWBNNTs, respectively.

## 5. Summary and conclusions

To summarize, an interlayer potential for *h*-BN and its derivative structures has been developed. The *h*-BN ILP was constructed from three main terms representing the interlayer attraction due to dispersive interactions, repulsion due to anisotropic overlap of electron clouds, and monopolar electrostatic interactions. The potential, which was parameterized against advanced dispersion-corrected DFT calculations, was shown to capture well both binding and lateral sliding energies of planar *h*-BN based dimer systems. Furthermore, the new interlayer potential was shown to reproduce DFT results of the interlayer sliding and rotation energy landscapes of curved DWBNNT of different crystallographic orientations. Our results indicate that when constructing force-fields aimed at modeling layered structures, their intrinsic anisotropic nature requires separate treatment of their intra- and inter-layer interactions. We note that the TS-vdW methods, which is used for the benchmark calculations, does not take into account screening effects, that may prove important for some systems.[65,66]

Nevertheless, our *h*-BN ILP is of general nature and can be readily parameterized against dispersion-corrected density functionals, that incorporate screening effects, when necessary. Therefore, we believe that the proposed force-field opens the way for accurate and efficient modeling and simulation of large-scale layered structures based on the promising material *h*-BN.

**Acknowledgments**:

Work at Tel-Aviv University was supported by the Israel Science Foundation under Grant No. 1313/08, the Center for Nanoscience and Nanotechnology at Tel Aviv University, the Israeli Ministry of Defense, and the Lise Meitner-Minerva Center for Computational Quantum Chemistry. Work at the Weizmann Institute of Science was supported by the European Research Council, the Israel Science Foundation, and the Lise Meitner-Minerva Center for Computational Quantum Chemistry.